\documentclass[aps,
prd,
preprint,
nofootinbib,
groupedaddress]{revtex4}

\usepackage{graphicx} 
\usepackage{amsmath} 
\usepackage{amssymb} 
\def\identity{{\rm 1\kern -.23em l}} 
\def\det{{\mathrm{det}}}

\def\eps{{\varepsilon}}
\def\Pp{{\frac{1+\gamma_5}{2}}}
\def\Pm{{\frac{1-\gamma_5}{2}}}
\newcommand{\beq}{\begin{equation}}
\newcommand{\eeq}{\end{equation}}
\newcommand{\bea}{\begin{eqnarray}}
\newcommand{\eea}{\end{eqnarray}}
\newcommand{\mo}{\mathcal{O}}

\begin{document}

\title{Anomalous fermion number nonconservation: Paradoxes in the level crossing picture}
\author{Y. Burnier} 
\email{Yannis.Burnier@epfl.ch}
\affiliation{
Institut de Th\'eorie des Ph\'enom\`enes Physiques,
Ecole Polytechnique F\'ed\'erale de Lausanne,
CH-1015 Lausanne, Switzerland}
\date{September 4, 2006}
\begin{abstract}
In theories with anomalous fermion number nonconservation, the level crossing picture is considered a faithful representation of the fermionic quantum number variation. 
It represents each created fermion by an energy level that crosses the zero-energy line from below. If several fermions of various masses are created, the level crossing picture contains several levels that cross the zero-energy line and cross each other. However, we know from quantum mechanics that the corresponding levels cannot cross if the different fermions are mixed via some interaction potential.
The simultaneous application of these two requirements on the level behavior leads to paradoxes.
For instance, a naive interpretation of the resulting level crossing picture gives rise to charge nonconservation. In this paper, we resolve this paradox by a precise calculation of the transition probability, and discuss what are the implications for the electroweak theory. In particular, the nonperturbative transition probability is higher if top quarks are present in the initial state.
\end{abstract}
%
\maketitle
\section{Introduction}
When a classical conservation law is broken by quantum corrections, It is said that the associated symmetry is anomalous. An anomaly in a current associated with gauge symmetry ruins the consistency of the theory. The requirement that all gauge anomalies cancel strongly restricts the possible physical theories. On the other hand, anomalies arising in other type of currents can lead to interesting physics. 
For instance in strong interactions, the anomaly in the chiral current is important in the well-known pion decay to two photons. In weak interactions, there is an anomaly in the baryon number current. Although anomalous baryon number violating transitions are strongly suppressed at small energies, they could be at the origin of the baryon asymmetry of the universe.

Anomalous transitions leading to fermion number nonconservation arise in the electroweak theory or any other model with a similar vacuum structure. The crucial feature is the existence of an infinite number of degenerate vacua, separated by energy barriers and the transition between them leads to the creation, or destruction, of fermions. The energy barrier can be passed by either by tunneling, which is represented by an instanton \cite{'tHooft:1976fv}, or by thermal excitations \cite{sphaleron,2Dmodel::sphaleron}. In the second case, the relevant configuration is the sphaleron \cite{Klinkhamer:1984di}. It is defined as the maximum energy configuration along the path of minimal energy connecting two neighboring vacua.

To visualize anomalous fermion number nonconservation, 
let us consider the path in the bosonic field space, parametrized by $\tau$, that relates two neighboring vacua via the sphaleron configuration. If the bosonic fields evolve very slowly along this path, the fermionic states can be found by solving the static Dirac equation $H_D\Psi_n=E_n\Psi_n$. This equation has positive as well as negative energy states. 
A way to represent the fermionic vacuum state is the Dirac sea. All states with negative energy are filled, whereas all positive energy states are empty. We are interested in the variation of the Dirac sea as a function of $\tau$. On a graph containing all energy levels as function of $\tau$, it may happen that an initially negative (therefore occupied) energy level crosses the zero energy line and becomes a real particle. This is the level crossing picture representation of the anomalous fermion number nonconservation \cite{Jackiw:1976pf}.
In the case of the electroweak theory, one level for each existing fermionic doublet crosses the zero-energy line in the transition between two adjacent vacua \cite{LCpicture}.

The level crossing picture can be thought of as a quantum mechanical description of fermion creation. This description is assumed to match the complete quantum field theory when the background evolves very slowly.

\begin{figure} 
\begin{center} 
\includegraphics[width=140mm,height=40mm]{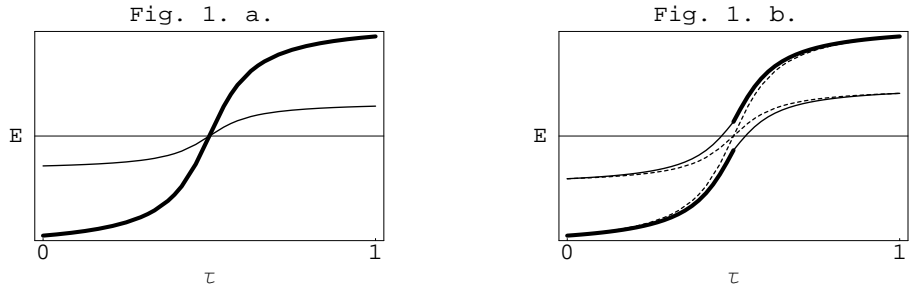} 
\caption{Naive picture of level crossing containing only the two levels that crosses the zero-energy line. The heavy particle is represented with a thick line. Two cases are pictured; without mixing (a) and with mixing between the two fermions (b). In the latter case the energy levels can't cross, therefore the heavy particle becomes a light one and vice-versa.} 
\label{LCnaif}
\end{center} 
\end{figure}

Consider now the case of two fermions $\Psi_i$, where $i=1,2$ is the flavor index. 
We first assume that the different flavors are not mixed by any interaction term, that is to say the Dirac equation, which generally reads $H_{ij}\Psi_j=E\Psi_j$, can be diagonalized in flavor space for any $\tau$. We will call fermions for which $H$ is diagonal, independent. On the level crossing picture for two fermions with different masses\footnote{We consider here fermions made massive through their Yukawa coupling to the Higgs field and not by a tree mass term.}, we see that two energy levels cross the zero-energy line and cross each other. A simplified level crossing picture containing these two levels is given in Fig. \ref{LCnaif}.a.\footnote{The full level crossing picture of the theory we consider in the following is given in Fig. \ref{LCsansint}.} 

On the other hand, if the two fermions are mixed, for instance by the interaction between them and the background sphaleron fields, and the Dirac Hamiltonian is not diagonalizable, we know from quantum mechanics that the energy levels cannot cross each other. Therefore, in this case, the heavy fermion becomes the light one and the light one becomes the heavy one, see Fig. \ref{LCnaif}.b. Indeed, generally there are excited states of the light particle and the light fermion evolves to one of them, see Fig. \ref{LCavecint}.
Therefore, we conclude from the level crossing picture that two light fermions are created in the case with mixing instead of a light and a heavy one.  

Suppose now that we introduce another gauge field $B_\mu$, which shall be Abelian, not spontaneously broken and free from any anomaly.  We further assume that the different fermions have different charges with respect to this field. The cancellation of the anomaly for this gauge field $B_\mu$ requires that the sum of all the fermionic charges vanishes and therefore an anomalous transition leading to the creation of one of each existing fermions perfectly respects the $B$-gauge symmetry. However, in the case of mixing between the fermions, the creation of only light fermions leads to $B$-charge violation.
 
Note that the mixing between the fermions is only possible if the background has a non-vanishing charge with respect to the $B$-gauge symmetry. It means that the $B$-gauge is broken in the sphaleron or instanton core. This is possible, and is generally the case, even if the $B$-gauge symmetry is unbroken in vacuum.

Two points deserve further investigations.  Firstly, if the level crossing picture changes qualitatively, it is interesting to see if the transition probability undergoes such changes. Secondly, we have to understand how charge conservation is ensured. 

These questions are mainly model independent,   
therefore we choose to resolve them in a simple 1+1 dimensional anomalous Abelian Higgs model with two chiral fermions, which contains the above paradox.  In this particular model, we will show that the probability for the creation of two light fermions is zero, unless it is accompanied by the emission of some other particles that compensate for the charge asymmetry. The level crossing picture must therefore be reinterpreted. 
We will also see that the nonperturbative transition is more probable if there are heavy fermions in the initial or final state, again in contradiction to what the level-crossing picture suggests.

These paradoxes also arise in the electroweak theory.  Indeed, the quarks have various electric charges and in the background of the electroweak sphaleron or instanton the $SU(2)\times U(1)$-gauge symmetry is completely broken by the presence of a charged weak field background.
The resolution of these questions is of great interest for electroweak baryogenesis\footnote{Even though with the current constraints on the Higgs mass, producing the observed baryonic asymmetry within the minimal Standard Model is impossible, electroweak baryogenesis may still work in some of its extensions \cite{Bodeker:2004ws} and in supersymmetric theories \cite{Carena:2004ha}.}.

This paper is organized as follows. In Sec. \ref{a4s2}, we design a 1+1 dimensional model adapted to our purposes. The level crossing picture of this model is derived in Sec. \ref{a4s3}. In Sec. \ref{a4s4}, we perform a computation of the non-perturbative transition rate in the instanton picture and resolve the paradoxes mentioned before. The implications of our results for the electroweak baryogenesis and a conclusion is given in Sec. \ref{a4s5}.

\section{The model} \label{a4s2}
We construct here a simple model which contains the paradoxes mentioned in the introduction.
A good candidate for studying the creation of massive fermions is 
the chiral Abelian Higgs model studied in \cite{2Dmodel::sphaleron,Bochkarev:1987wg,Burnier:2005he,Bezrukov:2005rw},
\begin{eqnarray} 
\mathcal{L} &=&-\frac{1}{4}F^{\mu \nu }F_{\mu \nu }+i\overline{\Psi 
}^j\gamma ^{\mu }(\partial _{\mu }-i\frac{e}{2}\gamma ^{5}A_{\mu 
})\Psi^j  \notag \\ &&-V(\phi )+\frac{1}{2}\left| D_{\mu }\phi 
\right| ^{2}+if^j\overline{\Psi}^j\frac{1+\gamma_5}{2}\Psi^j\phi^*  
-if^j\overline{\Psi}^j\frac{1-\gamma_5}{2}\Psi^j\phi,
 \label{La} 
\end{eqnarray} 
where $D_{\mu }\phi =\partial _{\mu }-ieA_{\mu }$, $V(\phi )=\frac{%
\lambda }{4}\left( \left| \phi \right| ^{2}-v^{2}\right) ^{2}$ and $\Psi^j=\binom{\Psi_L^j}{\Psi_R^j}$ with $j=1,2$ labeling the different flavors. We use the following representation for the $\gamma$-matrices:
\begin{equation}
\gamma _{0}=\left( \begin{array}{cc}
0 & -i \\ 
i & 0
\end{array}
\right) ,\text{ }\gamma _{1}=\left( 
\begin{array}{cc}
0 & i \\ 
i & 0
\end{array}
\right),\text{ }\gamma_5=\gamma_0\gamma_1 .
\end{equation}
This model displays similar nonperturbative properties as the electroweak theory.

We also need a mixing term between the different flavors, 
which would, at the semi-classical level, prevent the energy levels of the 
fermions from crossing each other. Interaction terms between fermions are not possible in this simple model. Indeed, if we write interaction terms between different fermions like $f_{ij}\Psi _{i,L}^{\dagger }\Psi _{j,R}\phi ^{\ast }+h.c.$ we could redefine the fields to diagonalize the matrix $f_{ij}$ and the theory would inevitably lead to independent fermions\footnote{Moreover the first scalar field in (\ref{La}) vanishes in the center of the instanton, which would allow
the levels to cross at the center.}. We therefore have to introduce another scalar field, allowing for other Yukawa couplings. We will also introduce a $U(1)^{B}$ gauge field $B_{\mu }$ to give different charges to the two different flavors.
The bosonic sector reads:
 \begin{eqnarray} 
 \mathcal{L} &=&-\frac{1}{4}F_{A\mu \nu }F_{A}^{\mu \nu }-\frac{1}{4}F_{B\mu 
 \nu }F_{B}^{\mu \nu }+\frac{1}{2}\left| (\partial _{\mu }-ieA_{\mu })\phi 
 \right| ^{2}+\frac{1}{2}\left| (\partial _{\mu }-ieA_{\mu }-ie'B_{\mu })\chi 
 \right| ^{2}  \notag \\ 
 &&-\frac{\lambda }{4}\left(\left| \phi \right| ^{2}-v^{2} \right) ^{2}-\frac{\Lambda }{4}\left| \chi \right| ^{4}-\frac{M^{2}}{2}%
 \left| \chi \right| ^{2}-\frac{h}{2}\left| \chi \right| ^{2}(\left| \phi 
 \right| ^{2}-v^{2}). \label{LbosonM}
 \end{eqnarray} 
%
%
%
We have now to specify the charge of each of the four spinor components $\Psi^{1,2}_{L,R}$ with respect to $U(1)^A$ and $U(1)^B$. Let us note $\alpha_{L,R}^{1,2}$ and $\beta_{L,R}^{1,2}$ the charges with respect to $A_\mu$ and $B_\mu$. The following choice turns out to serves our aim:
\begin{equation}
\begin{array}{cc}
\alpha _{R}^{1}=-\frac{e}{2}, & \alpha _{L}^{1}=\frac{e}{2}, \\ 
\alpha _{R}^{2}=-\frac{e}{2}, & \alpha _{L}^{2}=\frac{e}{2},
\end{array}
\text{ and }
\begin{array}{cc}
\beta _{R}^{1}=\frac{e'}{2}, & \beta _{L}^{1}=\frac{e'}{2}, \\ 
\beta _{R}^{2}=-\frac{e'}{2}, & \beta _{L}^{2}=-\frac{e'}{2}.
\end{array}
\text{ }
\end{equation}
The gauge symmetries imply that there are two classically conserved electric currents
\begin{eqnarray*}
j_{A}^{\mu } &=&\alpha _{L}^{i}\overline{\Psi }_{L}^{i}\gamma ^{\mu }\Psi
_{L}^{i}+\alpha _{R}^{i}\overline{\Psi }_{R}^{i}\gamma ^{\mu }\Psi _{R}^{i},
\\
j_{B}^{\mu } &=&\beta _{L}^{i}\overline{\Psi }_{L}^{i}\gamma ^{\mu }\Psi
_{L}^{i}+\beta _{R}^{i}\overline{\Psi }_{R}^{i}\gamma ^{\mu }\Psi _{R}^{i}.
\end{eqnarray*}
These currents are in general anomalous but are conserved with our particular choice of charges.
\begin{eqnarray*}
\partial _{\mu }j_{A}^{\mu } &=&\frac{1}{4\pi }\varepsilon _{\mu \nu
}F_{A}^{\mu \nu }\sum_{i}\left[ (\alpha _{R}^{i})^{2}-(\alpha _{L}^{i})^{2}%
\right] + 
\frac{1}{4\pi }\varepsilon _{\mu \nu }F_{B}^{\mu \nu }\sum_{i}\left[
\alpha _{R}^{i}\beta _{R}^{i}-\alpha _{L}^{i}\beta _{L}^{i}\right]  
=0, \\
\partial _{\mu }j_{B}^{\mu } &=&\frac{1}{4\pi }\varepsilon _{\mu \nu
}F_{B}^{\mu \nu }\sum_{i}\left[ (\beta _{R}^{i})^{2}-(\beta _{L}^{i})^{2}%
\right] + 
\frac{1}{4\pi }\varepsilon _{\mu \nu }F_{A}^{\mu \nu }\sum_{i}\left[
\alpha _{R}^{i}\beta _{R}^{i}-\alpha _{L}^{i}\beta _{L}^{i}\right]
=0.
\end{eqnarray*}
The fermionic current
\begin{equation}
j_{F}^{\mu }=\overline{\Psi }_{L}^{i}\gamma ^{\mu }\Psi _{L}^{i}+\overline{
\Psi }_{R}^{i}\gamma^{\mu }\Psi _{R}^{i}.
\end{equation}
is conserved at the classical level, however, its anomaly does not vanish:
\begin{eqnarray}
\partial _{\mu }j_{F}^{\mu } &=&\frac{1}{4\pi }\varepsilon _{\mu \nu
}F_{A}^{\mu \nu }\sum_{i}\left[ (\alpha _{R}^{i})-(\alpha _{L}^{i})\right] +
\frac{1}{4\pi }\varepsilon _{\mu \nu }F_{B}^{\mu \nu }\sum_{i}\left[
(\beta _{R}^{i})-(\beta _{L}^{i})\right]  \notag \\
&=&\frac{-e}{2\pi }\varepsilon _{\mu \nu }F_{A}^{\mu \nu }.
\end{eqnarray}
This is indeed what we need; there is no gauge anomaly but the fermion number current is anomalous.
We can now write down an interaction between fermions and scalar field, and the fermionic Lagrangian reads:
\begin{eqnarray}
\mathcal{L}_{ferm}^{Mink}=+i\overline{\Psi }^{1}\gamma ^{\mu }(\partial _{\mu }-e\frac{i}{2}%
\gamma_5 A_{\mu }-e'\frac{i}{2}B_{\mu })\Psi^{1}+i\overline{\Psi }^{2}\gamma
^{\mu }(\partial _{\mu }-e\frac{i}{2}\gamma_5 A_{\mu }+e'\frac{i}{2}B_{\mu })\Psi
^{2}  \notag \\
+if_{j}\overline{\Psi}^{j}\Pp\Psi^{j}\phi^* -if_{j}\overline{\Psi}^{j}\Pm\Psi^{j}\phi -if_3\overline{\Psi}^{1}\Pm\Psi^{2}\chi
+if_3\overline{\Psi}^{2}\Pp\Psi^{1}\chi^*  . \label{LfermionM}
\end{eqnarray}
The fermionic spectrum consists of two fermions of different mass $F^j=vf^j$, $j=1,2$ interacting with each other by Yukawa coupling to the scalar field $\chi$. 
The vacuum structure of the model given by the Lagrangians (\ref{LfermionM}) and  (\ref{LbosonM}) is non-trivial 
\cite{Jackiw:1976pf}. Taking the $A_0=0$ gauge and putting the theory 
in a spatial box of length $L$ with periodic boundary conditions, one 
finds that there is an infinity of degenerate vacuum states  
$|n\rangle,~ n \in \mathbf{Z}$ with the gauge-Higgs configurations 
given by 
\begin{equation} 
A_1= \frac{2\pi n}{eL},\quad \phi=ve^{i\frac{2\pi nx}{L}}. 
\label{vacst} 
\end{equation}   
The transition between two neighboring vacua   
leads to the creation of two fermions as intended: If the vector field $A_\mu$ undergoes the variation
\begin{equation}
\delta A_{1}=-\frac{2\pi }{Le},
\end{equation}
which corresponds to the difference between two adjacent vacua,
the fermionic anomaly is
\begin{equation}
\delta N_{F}=\int \frac{-e}{2\pi }\varepsilon ^{\mu \nu }F_{\mu \nu
}^{A}d^{2}x=\frac{-e}{2\pi }\int 2\partial _{0}A_{1}d^{2}x=\frac{-e}{2\pi }%
2\delta A_{1}L=2.
\end{equation}
\section{Level crossing picture}\label{a4s3}
We build a path in the bosonic field space that goes adiabatically from one vacuum to the neighboring one. To this aim, we find the sphaleron and construct a path that relates it with the initial and final vacua. Such configurations are relevant for high temperature dynamics \cite{Arnold:1987mh}.

Using  the $A_{0}=B_{0}=0$ gauge, the sphaleron in this model reads
\begin{eqnarray}
\phi ^{cl} &=& -ive^{-\frac{\pi ix}{L}}\tanh (Mx),\notag  \\
\chi ^{cl} &=&i\alpha e^{-\frac{\pi ix}{L}}\cosh ^{-1}(Mx),\notag \\
A_{1}^{cl} &=&-\frac{\pi }{eL}, \label{sphaleron} \\
B_{1}^{cl} &=&0,\notag
\end{eqnarray}
 with $\alpha =\sqrt{\frac{1}{h}(\lambda v^{2}-2M^{2})}$. 
It can be found using results on solitons with two scalar fields of Refs. \cite{Rajaraman:1978kd}.
Note that this solution is only valid for a restricted parameter space
\begin{eqnarray}
\lambda v^{2}>2M^{2},\\
2M^{2}+\Lambda \alpha ^{2}-hv^{2}=0.  \label{contrainte}
\end{eqnarray}

An example of a path going from vacuum $n=0$ at $\tau=0$ to vacuum $n=-1$ at $\tau=1$ via the sphaleron at $\tau=1/2$ is
\begin{eqnarray}
\phi ^{cl} &=&ve^{-\frac{2\pi ix\tau }{L}}\left[ \cos (\pi \tau )+i\sin (\pi
\tau )\tanh (Mx\sin (\pi \tau ))\right], \notag  \\
\chi ^{cl} &=&-i\alpha e^{-\frac{2\pi ix\tau }{L}}\sin (\pi \tau )\cosh ^{-1}(Mx\sin (\pi \tau)),\notag \\
A_{1}^{cl} &=&-\frac{2\pi \tau }{eL}  \label{2config}, \\
B_{1}^{cl} &=&0\notag.
\end{eqnarray}
These configurations represent a set of static background fields interpolating between vacua in which the fermions evolve. The equations of motion for the fermions are
\begin{equation}
i\partial_0\Psi=H\Psi\label{De},
\end{equation}
with
\begin{equation}
\Psi =\left( 
\begin{array}{c}
\Psi _{1}=\Psi _{L}^{1}\text{ } \\ 
\Psi _{2}=\Psi _{R}^{1} \\ 
\Psi _{3}=\Psi _{L}^{2} \\ 
\Psi _{4}=\Psi _{R}^{2}
\end{array}
\right)   \label{p1234}
\end{equation}
and
\begin{equation}
H=\left( 
\begin{array}{cccc}
{ -i\partial }_{1}{ -}\frac{e}{2}{ A}_{1}{ -}\frac{e'}{2}{ B}_{1} & 
\hspace{-0.6cm} { f}_{1}{ \phi } &\hspace{-0.6cm} { 0}  &\hspace{-0.6cm} { f}_{3}{ \chi }\\ 
{ f}_{1}{ \phi }^{\ast } &\hspace{-0.6cm} { i\partial }_{1}{ -}
\frac{e}{2}{ A}_{1}{ +}\frac{e'}{2}{ B}_{1} &\hspace{-0.6cm} { 0} &\hspace{-0.6cm} { 0} \\ 
{ 0} &\hspace{-0.6cm} { 0} &\hspace{-0.6cm} { -i\partial }_1
{ -}\frac{e}{2}{ A}_{1}{ +}\frac{e'}{2}{ B}_{1} &\hspace{-0.6cm} 
{ f}_{2}{ \phi } \\ 
{ f}_{3}{ \chi }^{\ast } &\hspace{-0.6cm}
{ 0} &\hspace{-0.6cm} { f}_{2}{ %
\phi }^{\ast } &\hspace{-0.6cm} { i\partial }_{1}{ -}\frac{e}{2}{ A}_{1}%
{ -}\frac{e'}{2}{ B}_{1}\label{DH}
\end{array}
\right) ,
\end{equation}
the Dirac Hamiltonian.
In the limit of slow transition $\dot\tau\sim 0$, the Hamiltonian is time-independent and the spectrum of the static Dirac equation $H\Psi_n=E_n\Psi_n$ for each $\tau$ leads to the level-crossing picture. Of course, an analytic solution to this eigenvalue problem is not possible for each $\tau$. We therefore give the analytic solutions at a few values of $\tau$, check with perturbation theory that the interaction potential lifts the degeneracy of the levels where they cross each other, and then give the complete level crossing picture resulting from numerical computation.   
\subsection{Fermionic spectrum in the vacuum $\tau =0,1$}
For $\tau=0,1$ the fermionic spectrum is the one of non-interacting fermions $\Psi^i,~i=1,2$ \cite{Bezrukov:2005rw} and is labeled by an integer $n$. 
\begin{eqnarray}
E_n^{i} &=&\pm\sqrt{F_{i}^{2}+k_n^{2}} \\
k_n &=&\left\{\begin{array}{cc} \frac{2\pi n}{L}, &\tau=0,\\
\frac{2\pi (n-\frac{1}{2})}{L}, &\tau=1.\end{array}\right.\notag
\end{eqnarray}
Note that the spectrum is different in the states $\tau=0$ and $\tau=1$. The configuration $\tau=0$ is not a true vacuum for fermions, the fermionic contribution to vacuum energy being larger for $\tau=0$ than for $\tau=1$. This difference however vanishes in the limit of infinite system size \cite{Bezrukov:2005rw}. All states are doubly degenerate in energy except for $\tau=0$ in the case $n=0$.
\subsection{Sphaleron configuration, $\tau =1/2$}
The Dirac Hamiltonian in the background of the sphaleron reads
\begin{equation}
{ H=}\left( 
\begin{array}{cccc}
{ -i\partial }_{1}{ +}\frac{\pi }{2L} &\hspace{-1cm} { iF}_{1}{ e%
}^{-i\frac{\pi x}{L}}\tanh(Mx) &\hspace{-1cm} 0 &\hspace{-1cm} { -iF}_{3}\cosh(Mx) ^{-1}{ e}^{-i\frac{\pi x}{L}}  \\ 
{ -iF}_{1}{ e}^{i\frac{\pi x}{L}}\tanh(Mx)  &\hspace{-1cm} { i\partial }_{1}%
{ +}\frac{\pi }{2L} &\hspace{-1cm} 0 &\hspace{-1cm} 0 \\ 
0 &\hspace{-1cm} 0 &\hspace{-1cm} { -i\partial }_{1}{ +}\frac{%
\pi }{2L} &\hspace{-1cm} { iF}_{2}{ e}^{-i\frac{\pi x}{L}}\tanh(Mx)  \\ 
{ iF}_{3}\cosh(Mx) ^{-1}{ e}^{i\frac{\pi x}{L}} &\hspace{-1cm} 0 &\hspace{-1cm} { -iF}_{2}{ e}^{i\frac{\pi x}{L%
}}\tanh(Mx)  &\hspace{-1cm} { i\partial }_{1} + \frac{\pi }{2L}
\end{array}
\right) .
\end{equation}
In the $F_3=\alpha f_3=0$ case, these equations decouple can be solved separately for each fermion. In the limit $~L\rightarrow \infty$, one finds two zero-modes, one for each fermion:
\begin{equation}
\Psi_0 ^{j}=\left( 
\begin{array}{c}
e^{-\frac{i\pi }{2L}x}[\cosh (Mx)]^{-\frac{F_{j}}{M}} \\ 
e^{\frac{i\pi }{2L}x}[\cosh (Mx)]^{-\frac{F_{j}}{M}}
\end{array}
\right),\quad j=1,2.  \label{psiinfiniLd1}
\end{equation}
The interaction can be introduced perturbatively. To this aim, the Dirac Hamiltonian is separated in two parts
$H =H_{0}+W$ with $H_{0}=H(f_{3}=0)$. In the
$\Psi_0^{1},~\Psi_0^{2}$ subspace, the interaction matrix reads
\begin{eqnarray*}
M_{ij} &=&\frac{1}{n_{i}n_{j}}\left\langle \Psi ^{i}\right| W\left| \Psi
^{j}\right\rangle  \\
&=&\left( 
\begin{array}{cc}
0 & iI \\ 
-iI & 0
\end{array}
\right),
\end{eqnarray*}
with
\begin{equation}
I=\frac{1}{n_{1}n_{2}}\int_{-L/2}^{L/2}(F_3)[\cosh (Mx)]^{-\frac{%
F_{1}+F_{2}}{M}-1}dx=\sqrt{\frac{\Gamma[\frac{F_1 + M}{2M}]\Gamma[\frac{F_2 + M}{2M}]}{\Gamma[\frac{F_1}{2M}] \Gamma[\frac{F_2}{2M}]}}\frac{ \Gamma[\frac{F_1 + F_2 + M}{2M}]}{\Gamma[1 +\frac{ F_1 + F_2}{2M}]},
\end{equation}
and $n_{i}=\left\langle \Psi^{i}\right. \left| \Psi^i\right\rangle ^{\frac{1}{2}}$.
The eigenstates of the matrix $M_{ij}$ are
\begin{eqnarray*}
\Psi _{+} &=&-i\Psi _{1}+\Psi _{2}\text{ with energy }E_{+}=I, \\
\Psi _{-} &=&\Psi _{1}-i\Psi _{2}\text{ with energy }E_{-}=-I.
\end{eqnarray*}
We see here that the interaction between the fermions lifts the degeneracy between
the states and avoids that the levels cross each other.
\subsection{Numerical results}
The energy levels may be found numerically for each value of $\tau$ solving the static Dirac equation with the Hamiltonian (\ref{DH}) and periodic boundary conditions in the interval of length $L$. 

\begin{figure} 
\begin{center} 
\includegraphics[width=120mm,height=71mm]{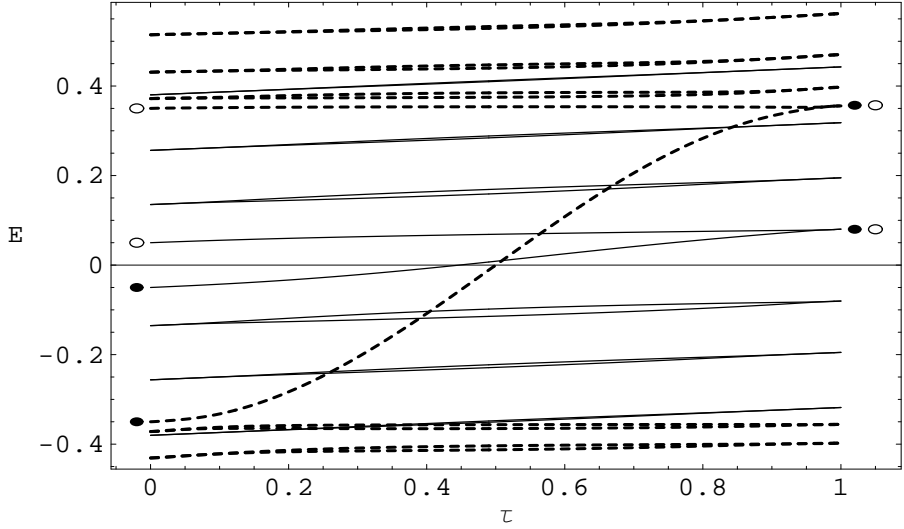} 
\caption{Level crossing of two fermions without mixing, $F_1=0.05,~F_2=0.35,~f_3=0,L=50$ and $h=m=e=e'=1,\;M=0.5$. One of each fermion is created when going from $\tau=0$ to $\tau=1$.} 
\label{LCsansint}
\end{center} 
\end{figure}

\begin{figure} 
\begin{center} 
\includegraphics[width=120mm,height=71mm]{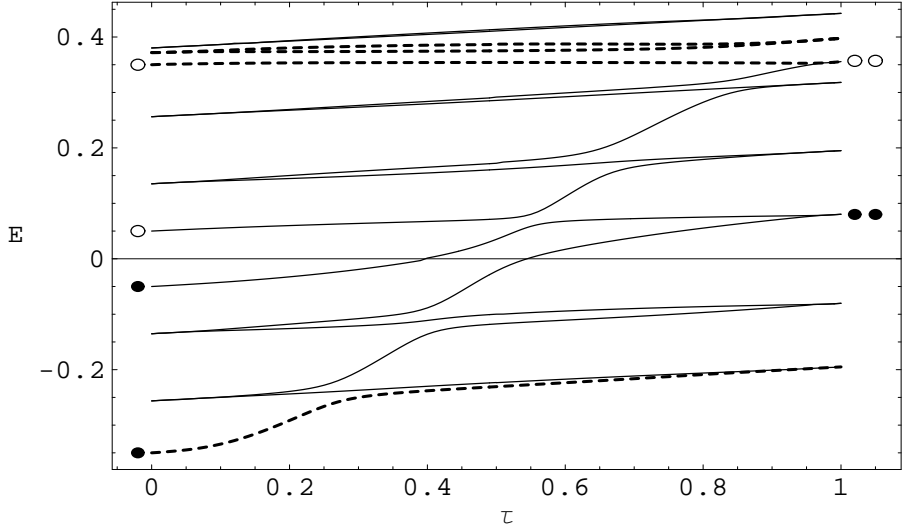} 
\caption{Level crossing of two fermions with mixing, $F_1=0.05,~F_2=0.35,~f_3=0.24,L=50$. Two light fermions are created here.} 
\label{LCavecint}
\end{center} 
\end{figure}

The results (Fig. \ref{LCsansint}, \ref{LCavecint}) show, in the cases of independent and mixed fermions, the creation of two fermions (two levels cross the zero-energy line). In the independent case, one of each fermion is created (Fig. \ref{LCsansint}), whereas two light ones are created in the mixed case (Fig. \ref{LCavecint}). The latter process violates charge conservation\footnote{Two light fermions of charge $-1/2$ with respect to the $B$ gauge field are created}. For charge conservation to be preserved, the transition probability of such a process must vanish. As a precise calculation of the transition probability is difficult in the sphaleron picture, we will use the instanton approach in the following, which leads to a well-defined semi-classical expansion. Note that the instanton picture will be similar to the adiabatic sphaleron transition if the fermionic masses are large and their associated time-scale small in comparison to the instanton size.

\section{Instanton picture}\label{a4s4}
We first derive the Euclidean properties of the model and then compute the transition probability for a few representative processes. 
In Euclidean space, the bosonic Lagrangian reads
 \begin{eqnarray} 
 \mathcal{L}_{bos}^{Eucl} &=&\frac{1}{4}F_{A\mu \nu }F_{A\mu \nu }+\frac{1}{4}F_{B\mu \nu 
 }F_{B\mu \nu }+\frac{1}{2}\left| (\partial _{\mu }-ieA_{\mu })\phi \right| 
 ^{2}+\frac{1}{2}\left| (\partial _{\mu }-ieA_{\mu }-ie'B_{\mu })\chi \right| 
 ^{2}  \notag \\ 
 &&+\frac{\lambda }{4}\left| \phi \right| ^{4}-\frac{m^{2}}{2}\left| \phi 
 \right| ^{2}+\frac{\Lambda }{4}\left| \chi \right| ^{4}+\frac{M^{2}}{2}%
 \left| \chi \right| ^{2}+\frac{h}{2}\left| \chi \right| ^{2}(\left| \phi 
 \right| ^{2}-v^{2}),  \label{LbosonE}
 \end{eqnarray} 
and the fermionic part
\begin{eqnarray}
\mathcal{L}_{ferm}^{Eucl}=+i\Psi^{\dagger 1}\gamma^{E}_{\mu }(\partial _{\mu }-e\frac{i}{2}%
\gamma_5 A_{\mu }-e'\frac{i}{2}B_{\mu })\Psi^{1}+i\Psi^{\dagger 2}\gamma^{E}_{\mu }(\partial _{\mu }-e\frac{i}{2}\gamma_5 A_{\mu }+e'\frac{i}{2}B_{\mu })\Psi
^{2}  \notag \\-if_{j}\overline{\Psi}^{j}\Pp\Psi^{j}\phi^* +if_{j}\overline{\Psi}^{j}\Pm\Psi^{j}\phi +if_3\overline{\Psi}^{1}\Pm\Psi^{2}\chi
-if_3\overline{\Psi}^{2}\Pp\Psi^{1}\chi^*. \label{LfermionE}
\end{eqnarray}
\subsection{Bosonic sector}
\begin{figure} 
\begin{center} 
\includegraphics[width=120mm,height=71mm]{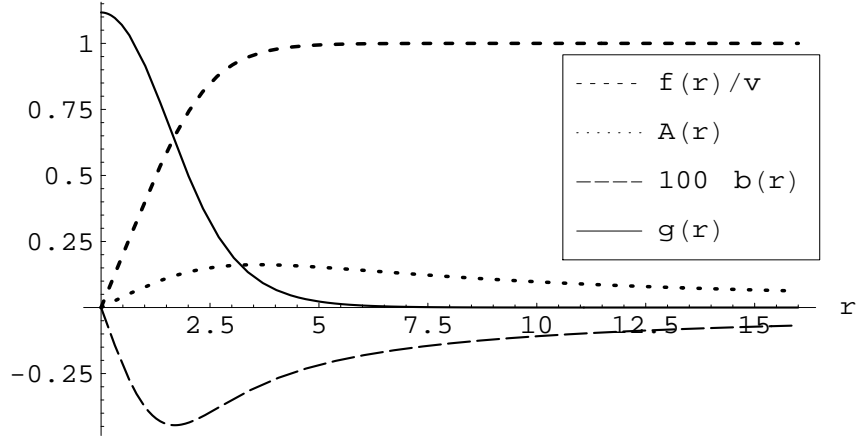} 
\caption{Instanton shape (with a different scale for the field $B$) for the following values for dimensionless constants (see Appendix \ref{AInst}): $m^2=\frac{M^2}{\lambda v^2} = 1, ~\mu =\frac{\lambda}{e^2}= 4,~\mu'=\frac{\lambda}{e'^2}= 4,~\rho =\frac{\Lambda}{h}= 1,~H =\frac{h}{\lambda}= 3$.} 
\label{instanton}
\end{center} 
\end{figure}

In order to find the instanton solution, let us point out the following: if 
$\chi =B=0$, we know the solution of the remaining equations, it is the 
 Nielsen-Olesen vortex \cite{Nielsen:1973cs}. We search here for a solution of the same type, adding some generic form for $B$ and $\chi$:
\begin{eqnarray}
\phi_{cl}(r,\theta )&=&f(r)e^{-i\theta},\notag\\
A_{cl}^{i}(r,\theta ) &=&\varepsilon ^{ij}\widehat{r}^{j}A(r)  \notag,\\
\chi_{cl}(r,\theta )&=&g(r),\label{inst_form}\\
B_{cl}^{i}(r,\theta )&=&\varepsilon ^{ij}\widehat{r}^{j}B(r)  \notag,
\end{eqnarray}
with polar coordinates $(it=\tau =r\cos \theta ,x=r\sin \theta )$, $\widehat{r}$ the unit vector in the direction of $r$ and $\varepsilon ^{ij} $ the completely antisymmetric tensor with $\varepsilon ^{01}=1$. 
Some details can be found in Appendix \ref{AInst}, only some results will be given here. An example of profile is given in Fig. \ref{instanton} and the asymptotic form of the different functions are
\begin{eqnarray}
&f(r)\overset{r\rightarrow 0}{\longrightarrow }f_0 r+\mo(r^3)
\label{l1},\quad &A(r)\overset{r\rightarrow 0}{\longrightarrow }a_0 r+\mo(r^3),\notag\\&
g(r)\overset{r\rightarrow 0}{\longrightarrow }g_0+\mathcal{O}(r^2),\quad &B(r)\overset{r\rightarrow 0}{\longrightarrow }b_0 r+ \mo(r^3),\\
& f(r)\overset{r\rightarrow \infty }{\longrightarrow }1+f_\infty K_0(\sqrt{2\lambda}v r),\label{l2}\quad &
A(r)\overset{r\rightarrow \infty }{\longrightarrow }\frac{1}{er}+a_\infty K_1(e v r),\notag\\ &g(r)\overset{r\rightarrow \infty }{\longrightarrow }g_\infty K_1(M r),\quad & B(r)\overset{r\rightarrow \infty }{\longrightarrow }\frac{b_\infty}{r},
\label{l4}
\end{eqnarray}
where $f_0,~ a_0,~ b_0,~ g_0,~f_\infty,~ g_\infty, ~a_\infty,~b_\infty$ are constants found by computing the exact instanton profile.
\subsection{Fermions}
The fermionic fluctuations in the background of the instanton (\ref{inst_form}) are given by $H\Psi=E\Psi$, with:
\begin{eqnarray}
H&=&\left( 
\begin{array}{cc}
H_1&I_2\\I_1&H_2
\end{array}
\right) \notag\\
H_1&=&\left( 
\begin{array}{cc}
\hspace{-3cm}-if_1\phi^*&\hspace{-3cm}i\partial_0-\partial_1+\frac{e}{2}\left(-A_0-iA_1\right)+\frac{e'}{2}\left(B_0+iB_1\right)\\
-i\partial_0-\partial_1+\frac{e}{2}\left(-A_0+iA_1\right)+\frac{e'}{2}\left(-B_0+iB_1\right)&if_1\phi
\end{array}
\right)\notag \\
H_2&=&\left( 
\begin{array}{cc}
\hspace{-3cm}-if_2\phi^*&\hspace{-3cm}i\partial_0-\partial_1++\frac{e}{2}\left(-A_0-iA_1\right)+\frac{e'}{2}\left(-B_0-iB_1\right)\\
-i\partial_0-\partial_1+\frac{e}{2}\left(-A_0+iA_1\right)+\frac{e'}{2}\left(B_0-iB_1\right)&if_2\phi
\end{array}
\right) \notag\\
I_1&=&\left( 
\begin{array}{cc}
-if_3\chi*&0\\0&0
\end{array}
\right),\quad I_2=\left( 
\begin{array}{cc}
0&0\\0&if_3\chi
\end{array}
\right).
\end{eqnarray}
The zero-modes are found solving the equations $H\Psi=0$, with H the Dirac operator in the background of the instanton. We use polar coordinates $(r,\theta)$ and expand fermionic fluctuations in partial waves $\Psi=\sum_{m=-\infty}^\infty e^{im\theta}\Psi_m$. This leads to the following equations:
\bea
\left(\frac{\partial}{\partial r}+\frac{m}{r}+\frac{e}{2}A(r)+\frac{e'}{2}B(r)\right)\Psi_m^1-f_1 f(r)\Psi_m^2-f_3 g(r)\Psi^4_{m-1}&=&0,\notag \\
\left(\frac{\partial}{\partial r}-\frac{m}{r}+\frac{e}{2}A(r)-\frac{e'}{2}B(r)\right)\Psi_m^2-f_1f(r)\Psi_m^1&=&0,\notag\\
\left(\frac{\partial}{\partial r}+\frac{m-1}{r}+\frac{e}{2}A(r)-\frac{e'}{2}B(r)\right)\Psi_{m-1}^3-f_2f(r)\Psi_{m-1}^4&=&0,\\
\left(\frac{\partial}{\partial r}-\frac{m-1}{r}+\frac{e}{2}A(r)+\frac{e'}{2}B(r)\right)\Psi_{m-1}^4-f_2f(r)\Psi_{m-1}^3-f_3 g(r)\Psi^1_{m}&=&0.\notag
\eea
In the case $f_3=0$ and $B(r)=0$, the two fermions decouple and their zero modes are \cite{Jackiw:1981ee}:
\begin{equation}
\psi^j(r)\propto\binom{1}{-1}\exp\left[-\int_0^r
dr'(vf_jf(r)+\frac{e}{2}A(r))\right],~~j=1,2.\label{mode0}
\end{equation}
If $f_3\neq 0$ the zero modes have the following asymptotics\footnote{We consider here the approximation $B=0$ (or $e'=0$), which does not lead to observable changes (see Fig. \ref{instanton})}
\bea
\psi^1_{cl}=\frac{\alpha_1}{\sqrt{r}}\left(\begin{array}{c}  e^{-F_1 r} \\ 
 -e^{-F_1 r}\\ -\beta_1 e^{-F_2 r} e^{-i\theta}\\ 
\beta_1 e^{-F_2 r}\left(1+\frac{1}{F_2r}\right)e^{-i\theta}\end{array}\right),\quad \psi^2_{cl}=\frac{\alpha_2}{\sqrt{r}}\left(\begin{array}{c} \beta_2  e^{-F_1r}\left(1+\frac{1}{F_1r}\right) e^{i\theta} \\ 
-\beta_2 e^{-F_1r} e^{i\theta}\\  -e^{-F_2r} \\ 
 e^{-F_2r} \end{array}\right),\label{mode0ferm}
\eea
where $\alpha_1$ and $\alpha_2$ are normalization constants and $\beta_1$ and $\beta_2$ vanishes in the decoupled $f_3=0$ case (\ref{mode0}). In the coupled system, the existence of the zero-modes can be checked using the method of Ref. \cite{Jackiw:1981ee}, but $\beta_1$ and $\beta_2$ have to be computed numerically. 
The coefficients $\beta_{1,2}$, which will be needed to compare transition probabilities, are given as a function of the fermion masses in Fig. \ref{beta1}.
Within some approximations, the constants $\beta_{1,2}$ can be found analytically (see Appendix \ref{smallf3}).

\begin{figure} 
\begin{center} 
\includegraphics[width=120mm,height=71mm]{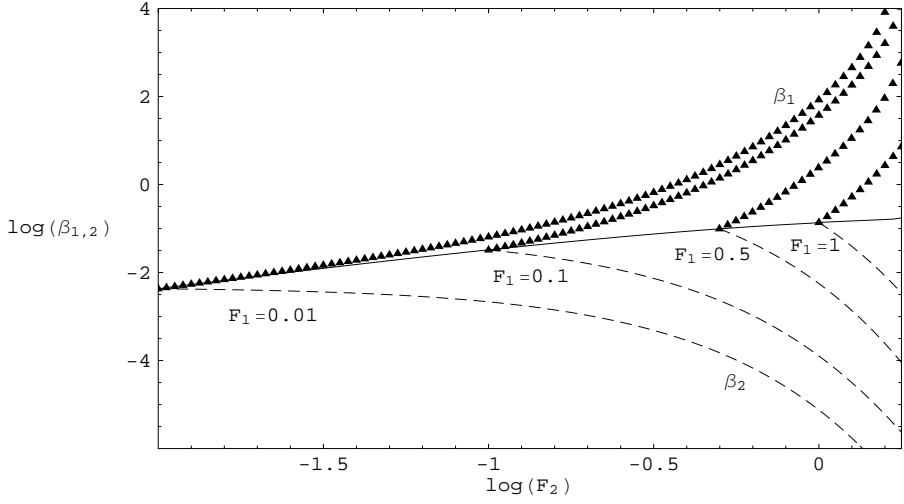} 
\caption{Coefficients $\beta_{2}$ (dashed lines) and $\beta_{1}$ (triangles) as a function of the mass $F_2$ for some different light fermion masses $F_1=0.01,~0.1,~0.5,~1$ and for $f_3=0.2$. The line represent $\beta_1=\beta_2$ in the degenerate case $F_1=F_2$. The constants $F_{1,2},~f_3$ are in units of $\sqrt{\lambda}v$.} 
\label{beta1}
\end{center} 
\end{figure}

\begin{figure} 
\begin{center} 
\includegraphics[width=130mm,height=71mm]{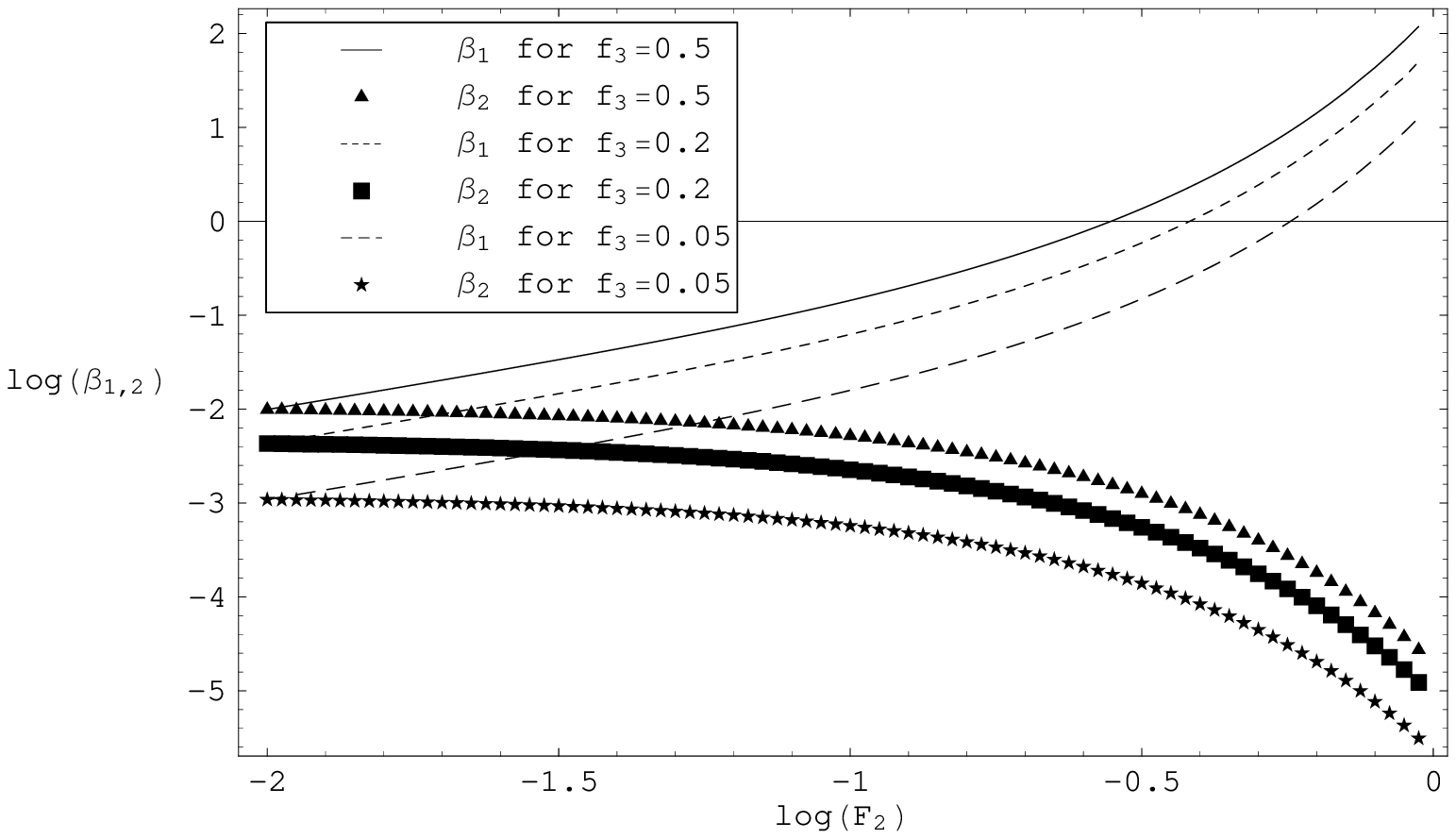} 
\caption{Coefficients $\beta_{1,2}$ as a function of the mass $F_2$ for some different couplings $f_3$ and for $f_1=0.01$. The constants $F_{1,2},~f_3$ are in units of $\sqrt{\lambda}v$.} 
\label{beta3}
\end{center}
\end{figure}

%
%
%
\subsection{Transition probability}
We start with two decoupled fermions ($f_3=0$) and introduce the interaction
perturbatively. It is clear what happens here; the interaction term $-if_3\overline{\Psi}^{1}\Pm\Psi^{2}\chi+h.c.$ allows for the decay of the heavy fermion into a light fermion and a $\chi$ boson, a process which conserves the charge and which can be taken into account in final state corrections (see \cite{finalstatecorr}, for inclusion of fermions see \cite{Espinosa:1991vq}). This is not what we we are interested in here. In the non-perturbative regime, two light fermions are created in the $\chi$ background, where the $U(1)^B$ gauge is broken and a $\chi$ boson should be emitted from the instanton tail as the $U(1)^B$ gauge symmetry is restored far from the instanton center. We will show here that processes violating charge conservation have vanishing probability.
\subsubsection{Green's function}
Green's functions with creation of two fermions and an arbitrary number of other particles read 
\bea
&&G^{ab}(x_1,x_2,y_1...y_n,z_1...z_m,w_1,...,w_l)=\\&& \int\mathcal{D}\Psi\mathcal{D}\overline{\Psi}\mathcal{D}\chi\mathcal{D}\chi^*\mathcal{D}\eta~ e^{-S[\Psi,\overline{\Psi},\chi,\chi*,\eta]}~\Psi^a (x_1)\Psi^b(x_2)\prod_{i=1}^n \chi(y_i)\prod_{j=1}^m \chi^*(y_j)\prod_{k=1}^l \eta(y_k),\notag
\eea
where $\eta$ stands for all neutral bosonic degrees of freedom, $\prod_{k''=1}^l \eta(y_k'')$ may contain the field $A,\phi$ and neutral pairs of fermions and the variable $\Psi$ is a spinor containing the two fermions as in (\ref{p1234}) and $a,b=1,..4$. 

The main contribution to the Green's function for the creation of two fermions comes from the sector with one instanton $(q=-1)$\footnote{More precisely, in the dilute instanton gas approximation the result from the one instanton sector can be exponentiated to give the complete Green's function \cite{coleman}.}. In this sector, fermions have two zero-modes (\ref{mode0ferm}). The Gaussian path integral over fermionic degrees of freedom can be evaluated, leading to the fermionic determinant with zero-modes excluded and the product of the fermionic zero-mode wave functions\footnote{Note that the zero-mode functions still depend on the background $\psi^i=\psi^i[\chi,\chi^*,\eta]$.},
\bea
&&G(x_1,x_2,y_1...y_n,z_1...z_m,w_1,...,w_l)= \int_{q=-1}\hspace{-4mm}\mathcal{D}\chi\mathcal{D}\chi^*\mathcal{D}\eta\\&&\times e^{-S[\Psi,\overline{\Psi},\chi,\chi*,\eta]}~\det'(K[\chi,\eta])\psi^1_{\chi,\eta}(x_1) \psi^2_{\chi,\eta}(x_2)\prod_{i=1}^n \chi(y_i)\prod_{j=1}^m \chi^*(y_j)\prod_{k=1}^l \eta(y_k).\notag
\eea
\subsubsection{Collective coordinates in the one instanton sector}
The bosonic action is expanded around the instanton configuration. Gaussian integration over the quadratic fluctuations gives a determinant $\det(D^2_{bos})^{-\frac{1}{2}}$. 
However, zero modes associated to symmetries require introduction of collective coordinates. There are two translation zero modes and one coming from $U(1)^B$ gauge. Performing an infinitesimal global gauge transformation, we get
\beq
\delta\chi=e^{i\beta}\chi-\chi\sim i\beta e^{i\theta}g(r),\quad \delta\phi=\delta A=\delta B=0. 
\eeq
Note that the $U(1)^A$ gauge is broken, there is no normalizable zero mode associated to this symmetry.
Rotation symmetry does not lead to a further zero mode.\footnote{Rotations give the same zero-mode as $U(1)^B$ gauge transformations}

Collective coordinates are introduced as follows. The integral over the translation zero modes are replaced by an integral over the instanton position $x_0$. The integral over the $U(1)^B$ gauge zero-mode is replaced by an integral over all possible global gauge transformations $\beta$. The Green's function reads:
\bea
&&G(x_1,x_2,y_1...y_n,z_1...z_m,w_1,...,w_l)= \int d^2x_0~d\beta~ e^{-S_{cl}}~\det'(K_{inst})N_B N_{tr}\\&& \times\det'(D^2_{bos})^{-\frac{1}{2}}\psi_\beta^1(x_1-x_0) \psi_\beta^2(x_2-x_0)\prod_{i=1}^n \chi_\beta(y_i-x_0)\prod_{j=1}^m \chi_\beta^*(y_j-x_0)\prod_{k=1}^l \eta_\beta(y_k-x_0)\notag,
\eea
with 
\bea
\chi_\beta=e^{i\beta}\chi_{cl}, \quad \chi^*_\beta=e^{-i\beta}\chi_{cl},\\
\eta_\beta=\eta_{cl},\quad \psi_\beta^j=e^{i\frac{\beta}{2}\Gamma_5}\psi^j_{cl},~j=1,2,
\eea
and $N_B,N_{tr}$ the normalization factor coming from variable change to collective coordinates.
To simplify the notations, we also introduced the matrices $\Gamma_{i},~i=1,2,5$ acting on the four dimensional spinor (\ref{p1234}) as:
\beq
\Gamma_1=\left(\begin{array}{cc}
\identity_2&0\\0&0\end{array}\right),\quad
\Gamma_2=\left(\begin{array}{cc}
0&0\\0&\identity_2\end{array}\right),\quad
\Gamma_5=\left(\begin{array}{cc}
\identity_2&0\\0&-\identity_2\end{array}\right),
\eeq
where $\identity_2$ is the identity on a two dimensional subspace.
\subsubsection{Fourier transformation of the Green's function}
The Fourier transformation of the Green's function after integration over the instanton location $x_0$ reads (writing spinor indices explicitly)
\bea
&&G^{ab}(k_1,k_2,p_1...p_n,p'_1...p'_m,q_1,...,q_l)= (2\pi)^2\delta^{(2)}\left(P\right)\int_0^{2\pi} d\beta~   \kappa\\&&\times\left(e^{i\frac{\beta}{2}\Gamma_5}\tilde{\psi}_{cl}^1(k_1)\right)^a\left(e^{i\frac{\beta}{2}\Gamma_5} \tilde{\psi}_{cl}^2(k_2)\right)^b\prod_{i=1}^n e^{i\beta} \tilde{\chi}_{cl}(p_i)\prod_{j=1}^me^{-i\beta} \tilde{\chi}_{cl}^*(p'_j)\prod_{k=1}^l \tilde{\eta}_{cl}(q_k)\notag,
\eea
where $\kappa=e^{-S_{cl}}~\det'(K_{inst})N_B N_{tr}\det(D^2_{bos})^{-\frac{1}{2}}$ and $P=k_1+k_2+\sum_{i=1}^n p_i+\sum_{i=1}^m p'_i+\sum_{i=1}^l q_i$. The integration over the instanton location leads to momentum conservation. In a similar way, integration over gauge rotation $\beta$ enforces charge conservation. Indeed, the integral over $\beta$ is non-zero only if the powers of $e^{i\beta}$ cancel, that is to say, if charge with respect to the gauge field $B_\mu$ is conserved.\footnote{Note that this do not depend on the existence of the $B_\mu$ field, but on the existence of the associated global symmetry. Therefore the requirement of charge conservation will persist in the limit $e'\to0$.}

As the different components of the spinors have different powers of $e^{i\beta}$, different cases have to be considered. We will concentrate here on three interesting situations, from which we will be able to derive some general conclusions. 
\subsection{Examples of allowed matrix elements}
First consider a process involving one $\phi$ scalar as initial state, which decays into two fermions. In this case the integration over the coordinate $\beta$ leads to:
\bea
&&G^{ab}(k_1,k_2,q_1
)= (2\pi)^2\delta^{(2)}\left(P\right) \kappa\tilde{\phi}_{cl}(q_1)
\\&&\times\left(\left(\Gamma_1\tilde{\psi}_{cl}^1(k_1)\right)^a\left(\Gamma_2 \tilde{\psi}_{cl}^2(k_2)\right)^b+ \left(\Gamma_2\tilde{\psi}_{cl}^1(k_1)\right)^a\left(\Gamma_1 \tilde{\psi}_{cl}^2(k_2)\right)^b \right) \notag.
\eea
Applying the reduction formula, we get a non-vanishing matrix element for two different fermions only by multiplying the Green's function by two fermionic legs $\bar u^1(k_1),~\bar u^2(k_1)$,
\bea
iM(k1,k2,q1)=(2\pi)^2\delta^{(2)}\left(q-k_1-k_2\right) \kappa~i(q^2+m_H^2)\tilde{\phi}_{cl}(q)\\ \left(\notag ~i\bar u^1 (k_1)(\hat{k_1}+F)\Gamma_1\psi_{cl}^1(k_1)~i\bar u^2(k_2)(\hat{k_2}+F)\Gamma_1\psi_{cl}^2(k_2)\right).
\eea
A straightforward calculation gives (see Appendix \ref{TF})
\bea
|M(k1,k2,q1)|^2=\left(2(2\pi)^3\kappa f_\infty\alpha_1\alpha_2 \left(1+\beta_1\beta_2\right)\right)^2.
\eea
The decay rate is after integration of the phase space (supposing $m_1<<m_\chi$):
\beq
\Gamma_\phi=\frac{1}{2m_\phi}\int d\mathrm{Lips } |M(k1,k2,q1)|^2=\frac{1}{2m_\phi(m_\phi^2- F_2^2)}(2(2\pi)^3\kappa f_\infty \alpha_1\alpha_2(1+\beta_1\beta_2))^2.\label{42}
\eeq

Secondly, we consider a process involving one $\chi$ scalar as initial state. The Fourier transformation of the Green's function reads
\bea
&&G^{ab}(k_1,k_2,q)= (2\pi)^2\delta^{(2)}\left(q-k_1-k_2\right) \kappa~\tilde{\chi}^*_{cl}(q)\left(\Gamma_1\tilde{\psi}_{cl}^1(k_1)\right)^a\left(\Gamma_1 \tilde{\psi}_{cl}^2(k_2)\right)^b \notag.
\eea
Applying the reduction formula, we get the matrix element for the creation of two light fermion by multiplying the Green's function by two light fermion legs $\bar u^1(k_1),~\bar u^1(k_2)$,
\bea
iM(k1,k2,q1)=(2\pi)^2\delta^{(2)}\left(q-k_1-k_2\right) \kappa~i(q^2+m_\chi^2)\tilde{\chi}^*_{cl}(q)\\ \notag ~i\bar u^1 (k_1)(\hat{k_1}+F)\Gamma_1\psi_{cl}^1(k_1)~i\bar u^1(k_2)(\hat{k_2}+F)\Gamma_1\psi_{cl}^2(k_2).
\eea
A straightforward calculation gives (see Appendix \ref{TF})
\bea
|M(k1,k2,q1)|^2=(2(2\pi)^3\kappa g_\infty \alpha_1\alpha_2\beta_2)^2.
\eea
The decay rate after integration of the phase space (supposing $m_1<<m_\chi$) reads
\beq
\Gamma_\chi=\frac{(2(2\pi)^3\kappa g_\infty \alpha_1\alpha_2\beta_2)^2}{2m_\chi^2\sqrt{m_\chi^2-4m_1^2}}.\label{45}
\eeq
A similar process involves the scalar $\chi^*$, which decays into two heavy fermions:
\beq
\Gamma_\chi=\frac{(2(2\pi)^3\kappa g_\infty \alpha_1\alpha_2\beta_1)^2}{2m_\chi^2\sqrt{m_\chi^2-4m_2^2}}.\label{46}
\eeq

Generalizing these three examples, we see that any interaction leading to the creation of two light respectively heavy fermions contains a factor $\beta_2$ respectively $\beta_1$. Processes leading to the creation of one of each fermion contain a factor $1+\beta_1\beta_2\sim 1$.
\subsection{Discussion of the different transition probabilities}
The integration over the collective coordinate associated with the gauge symmetry $U(1)^B$ leads necessarily to charge conservation. Therefore the process described by the level crossing picture (Fig. \ref{LCavecint}) cannot take place without the emission of some other particle that compensates the additional $U(1)^B$ charge. The possible initial and final states are more restricted than suggested by the level crossing picture.

We shall now discuss the transition probability of allowed processes. We leave aside for the moment the phase space factors, they are not large in 1+1 dimensions. The main factors that distinguish the transition rates (\ref{42}), (\ref{45}), (\ref{46}) for the three possible fermionic final states are the constants $\beta_{1,2}$. This is also true if more complicated processes are considered. As expected, if the fermions are light and weakly coupled, the probability to create one of each fermion is much larger; it is proportional to $1+\beta_1\beta_2\cong 1$, see Fig. \ref{beta1}, \ref{beta3}. However, in the case where one fermion is very heavy, it is unexpectedly favored to create two heavy fermions. The creation of two light fermions suggested by the level crossing picture is indeed suppressed. In the case of slow transitions (heavy fermion masses $F_{1,2}$ or large instanton radius $r_{inst}$) the probability of creating light fermions is larger, but it is still suppressed if the mass hierarchy is large ($\frac{F_2}{F_1}\gg 1$). 
\section{Conclusion}\label{a4s5}
In the model considered here, the level crossing picture suggests a particular transition which must not and does not occur. A possible way out would be to reinterpret it as follows. The level crossing picture only knows about fermions and the correct bosonic content of the initial and final states should be added by hand when dealing with a physical transition. More precisely, all symmetries that are broken by the fermionic initial and final states should be restored by supplementary bosonic operators. However, even with this extra requirement, the level crossing picture suggests the creation of two light fermions, a transition that turns out to be suppressed. Furthermore the most probable transition, computed with the inclusion of the first quantum corrections, would impose the energy levels to cross each other several times on the level crossing picture in spite of the interaction potential. Note that this is perfectly possible in quantum field theory although forbidden in the adiabatic quantum mechanical description.

The results for the transition probability are rather surprising; for heavy fermions, such as the top quark, or adiabatic process $r_{inst}F_2\gg 1$ (Sphaleron at high temperature) the probability of creating two heavy fermions is large. 
In the realistic electroweak theory, the phase space factor may be dominant and may change this conclusion. It is therefore very interesting to reproduce similar computations in the frame of the electroweak theory at high temperature, or at high energies.

A more interesting setup would be to include heavy quarks in the initial states. The phase space factor as well as the matrix element are then large. In this case, the nonperturbative transition rate can be enhanced by a huge factor (see Fig. \ref{beta1}). A high top quark density could therefore catalyze the nonperturbative transition rate. This phenomenon is relevant for baryogenesis at the electroweak phase transition. It could provide a mechanism to enhance the baryon number violating transition rate in the symmetric phase, while suppressing it in the broken phase. Indeed, while bubbles of true asymmetric vacuum expand in the symmetric universe, it may be that top quarks are more reflected by the bubble wall and are rare inside the bubble, and over-dense outside. This density asymmetry will render the nonperturbative rate faster outside the bubble, while slower inside.

It should be noted that the present calculation deals with the instanton rate, although at high temperature, the sphaleron rate is the relevant quantity. It would therefore be very interesting to find out if the sphaleron rate also displays these interesting features.
\section{Acknowledgments}
I am grateful to M. Shaposhnikov for drawing my attention to this problem as well as for many useful suggestions. I also thank F. Bezrukov and S. Khlebnikov for helpful discussions, and M. Weiss for comments on the manuscript. This work has been supported by the Swiss Science Foundation.
\appendix
\section{The instanton}\label{AInst}
 The bosonic Lagrangian (\ref{LbosonE}) in the $\partial _{\mu 
 }A_{\mu }=\partial _{\mu }B_{\mu }=0$ gauge gives the following equations of motion,
 \begin{eqnarray} 
 -\partial _{\mu }\partial _{\mu }\phi +2ieA_{\mu }\partial _{\mu }\phi 
 +e^{2}A_{\mu }^{2}\phi -\lambda v^{2}\phi +\lambda \left| \phi \right| ^{2}\phi 
 +h\left| \chi \right| ^{2}\phi  &=&0  \label{LdequaC1} , \notag\\ 
 -\partial _{\mu }\partial _{\mu }\chi +2i(eA_{\mu }+e'B_{\mu })\partial _{\mu 
 }\chi +(eA_{\mu }+e'B_{\mu })^{2}\chi ,\notag\\ +M^{2}\chi 
 +\Lambda \left| \chi \right| ^{2}\chi +h(\left| \phi \right| ^{2}-v^{2})\chi 
 &=&0  \label{LdequaC2} ,\notag \\ 
 -\partial _{\upsilon }\partial _{\upsilon }A_{\mu }+i\frac{e}{2}(\phi ^{\ast 
 }\overleftrightarrow{\partial }_{\mu }\phi +\chi ^{\ast }\overleftrightarrow{%
 \partial }_{\mu }\chi )+e^{2}A_{\mu }(\left| \phi \right| ^{2}+\left| \chi \right| ^{2})+ee'B_{\mu }\left| \chi \right| ^{2} &=&0, \\ 
 -\partial _{\upsilon }\partial _{\upsilon }B_{\mu }+i\frac{e'}{2}\chi ^{\ast }%
 \overleftrightarrow{\partial }_{\mu }\chi +e'^{2}B_{\mu }\left| \chi \right| ^{2}+e'eA_{\mu }\left| \chi \right| ^{2} &=&0. \notag
 \end{eqnarray} 
We are looking for a solution of the type (\ref{inst_form}).
As for the Nielsen-Olesen vortex, we impose the asymptotic behavior of the functions $A$ and $f$: 
\begin{eqnarray}
&&f(r)\overset{r\rightarrow 0}{\longrightarrow }f_1 r
, \quad
f(r)\overset{r\rightarrow \infty }{\longrightarrow }1,\label{l2a} \quad
A(r)\overset{r\rightarrow 0}{\longrightarrow }a_1 r\quad
A(r)\overset{r\rightarrow \infty }{\longrightarrow }\frac{1}{er}.
\end{eqnarray}
For the finiteness of the action, the function g(r), B(r) should respect the following boundary conditions:
\begin{eqnarray}
B(r)\overset{r\rightarrow 0}{\longrightarrow } b_1 r,\quad
g(r)\overset{r\rightarrow 0}{\longrightarrow }g_0,\quad B(r)+rB'(r)\overset{r\rightarrow \infty }{\longrightarrow }0,\quad g(r)\overset{r\rightarrow \infty }{\longrightarrow }0.
\end{eqnarray}
We also introduce dimensionless variables with the substitutions
\begin{eqnarray}
A=\tilde{A}\frac{\sqrt{\lambda v^2}}{e}, \quad f=\tilde{f}\frac{\sqrt{\lambda v^2}}{e},\quad
g= \sqrt{\frac{\lambda}{h}}\tilde{g}, \quad B=\tilde{B}\frac{\sqrt{\lambda v^2}}{e'},\quad
r=\frac{\tilde{r}}{\sqrt{\lambda v^2}}.
\end{eqnarray}
The remaining parameters are
\begin{eqnarray}
\mu=\frac{\lambda}{e^2},\quad \mu'=\frac{\lambda}{e'^2}, \quad \rho=\frac{\Lambda}{h},\quad
H=\frac{h}{\lambda}, \quad m^2=\frac{M^2}{\lambda v^2}.
\end{eqnarray}
The equations of motion (\ref{l2a}) in polar coordinates and with the ansatz (\ref{inst_form}) reads
\begin{eqnarray}
-\tilde{f}''(r)-\frac{1}{\tilde{r}}\tilde{f}'(r)+\left[\tilde{A}(r)-\frac{1}{\tilde{r}}\right]^2\tilde{f}(r)\notag\\
+\mu \tilde{f}(r)^3-\tilde{f}(r)+\tilde{g}(r)^2\tilde{f}(r)&=&0,\notag\\
-\tilde{g}''(r)-\frac{1}{\tilde{r}}\tilde{g}'(r)+(\tilde{A}(r)+\tilde{B}(r))^2\tilde{g}(r)+\rho \tilde{g}(r)^3\notag\\+m^2\tilde{g}(r)+H \tilde{g}(r)(\mu\tilde{f}(r)^2-1)&=&0,\notag\\
-\tilde{A}''(r)-\frac{\tilde{A}'(r)}{\tilde{r}}+\frac{\tilde{A}(\tilde{r})}{\tilde{r}^2}+\tilde{f}(r)^2\left[\tilde{A}(r)-\frac{1}{\tilde{r}}\right]\notag\\+\frac{1}{H\mu}\left(\tilde{B}(r)+\tilde{A}(r)\right)\tilde{g}^2(r)&=&0,\\
-\tilde{B}''(r)-\frac{\tilde{B}'(r)}{\tilde{r}}+\frac{\tilde{B}(r)}{\tilde{r}^2}+\frac{1}{H\mu'}\left(\tilde{B}(r)+\tilde{A}(r)\right)\tilde{g}^2(r)&=&0\notag,
\end{eqnarray}
where the prime means derivative with respect to $\tilde{r}$.
\section{Analytical approximations for the fermionic zero-modes}\label{smallf3}
For small coupling $f_3$, and small instanton size $a$ in units of fermion mass, we can get a rough approximation by perturbation theory. We checked numerically, that it corresponds reasonably well to the exact case and will be sufficient for the following discussion. We are interested in the case were the first fermion is very light in comparison to the second one and in comparison to the scalar field, $F_1\ll m_\chi$.

The first step is to eliminate the field $A_\mu$ by the variable change $\Psi\to \exp\left(-\frac{e}{2}\int dr A(r)\right)\Psi$ and contract the four first order differential equations into two second order ones. One obtains the following equations for the new variable $\Psi$:
\bea \left(-\frac{m(m-1)}{r^2}+\frac{f'(r) m}{r
   f(r)}-f(r)^2 f_1^2\right)\Psi _2(r)-\frac{f'(r) \Psi _2'(r)}{f(r)}+\Psi _2''(r)\notag\\=f(r) g(r) f_1 f_3 \Psi _4(r),\\
\left(-\frac{m(m-1)}{r^2}-\frac{f'(r) (m-1)}{r
   f(r)}-f(r)^2 f_2^2\right)\Psi _3(r) -\frac{f'(r) \Psi
   _3'(r)}{f(r)}+\Psi _3''(r)\notag\\=f(r) g(r) f_2 f_3 \Psi _1(r),\label{psi3}
\eea
with
\beq
\Psi _1(r)=\frac{1}{f(r) f_1}\left(\Psi _2'(r)-\frac{m \Psi _2(r)}{r}\right),\quad\Psi _4(r)=\frac{1}{f(r) f_2}\left(\frac{(m-1) \Psi _3(r)}{r}+\Psi _3'(r)\right).
\eeq
We discuss the case of the zero-mode $\psi^1_{cl}$ in $m=0$ partial wave first. At zero-order of perturbation we have the two first components ($\psi^1_{1,2}$) given by (\ref{mode0}) and the two last ones ($\psi^1_{3,4}$) vanish. If we consider now a non-vanishing $f_3 g(r)$ in (\ref{psi3}), the function $\psi^1_{3}$ is given at first order perturbation theory by
\beq
\Psi_3(r)=f_2 f_3\int dr' G(r,r') f(r') g(r') \psi^1 _1(r'),\label{intg}
\eeq
where $G(r,r')$ is the Green's function of the differential operator in the left hand side of equation (\ref{psi3}). We were not able to find a general expression for $G(r,r')$ for an arbitrary function $f(r)$, but satisfactory results are obtained using the Green's function $G(r,r')$ for constant\footnote{This approximation is exact in the limit of small instanton size and precise for light fermions, because they do not probe the instanton center.} $f(r)=v$. In this particular case
\beq
G(r,r')=-\frac{1}{f_2}\sinh(f_2r_<)\exp(-f_2r_>),\label{g}
\eeq
with $r_>=\max(r,r')$, $r_<=\min(r,r')$. Form (\ref{intg}), we get, for $r\gg 1$:
\beq
\Psi_3(r)\cong - f_3 v \exp(-f_2 r)\int dr'\sinh(f_2 r') g(r') \Psi _1(r'),\label{intg1}
\eeq
If the inverse fermion mass $\frac{1}{f_1v}$ is small in comparison to the typical extent $r_{inst}$ of the function $g(r)$, we have:
\bea
\beta_1=f_3 v \int_0^\infty d x g(x) \sinh(f_2 x) e^{-f_1 x} \sim f_3 v \int_0^\infty d x g(x) \sinh(f_2 x)
\eea
In the case of a large mass $f_2$ and large instanton size, the constant $\beta_1$ can be large from the presence of the sinh. In the case of small $f_2$, the integral can be further simplified to $\beta_1=f_3 F_2 \int x g(x) dx$.
A similar computation\footnote{The relevant green's function is given by (\ref{g}), were we replaced $f_1\to f_2$.} can be performed for $\psi_2^{cl}$,
\bea
\beta_2=f_3 v \int  dx \;g(x)\sinh(f_1 x)e^{-f_2 x}\label{beta2}.
\eea
If $f_1 r_{inst}\ll 1$ we have $\beta_2=f_3 F_1 \int  dx x g(x)e^{-f_2 x}$, which may be large for a large instanton size, if it is not suppressed by a large fermion mass $f_2$.
\section{Fourier transforms}\label{TF}
For computing cross sections, the unitary gauge is best suited. It is however known to be singular, which may lead to discontinuities in the fermionic wave functions. This can be easily cured using the following regularized gauge condition:
\beq
\alpha(r,\theta)=\theta-2\pi\Theta^\eps(\theta-\pi),
\eeq
where $\Theta^\eps(\theta-\pi)$ is continuous and goes to the step function as $\eps\to 0$ (see Ref. \cite{Bezrukov:2005rw} fore more details).
The Fourier transforms\footnote{We retain only the pole term here, the rest do not contribute to the final amplitude \cite{Kripfganz:1989vm}} of the fields are (we consider here only the case $f_3=0$ for fermions):
\bea
&&{\tilde\phi}(p)=\frac{f_\infty}{p^2+m_H^2},\quad \tilde{\chi}(p)=\frac{g_\infty}{p^2+m_\chi^2}\frac{pe^{i\theta_p}}{m},\label{ft}\\&&\tilde{A}_\mu(p)=\frac{ia_\infty}{m_W}\frac{\varepsilon_{\mu\nu}p_\nu}{m_W^2+p^2},\quad \psi^j_{R,L}(p) = -ic_\infty\sqrt{\frac{2}{\pi p}}e^{\frac{i}{2}\gamma_5\theta_p}\frac{F_j+p}{F_j^2+p^2},\notag
\eea
where $p=\sqrt{p_\mu p_\mu}$ and $\theta_q$ the angle between the spacial axis and the vector $p$. If $f_3\neq0$ but the field $B$ is neglected, the Fourier transforms of the two fermionic zero modes read:
\bea
\tilde{\psi}^1_{cl}(p) = \left(\begin{array}{c}
-i\alpha_1\sqrt{\frac{2}{\pi p}}e^{\frac{i}{2}\theta_p}\frac{F_1+p}{F_1^2+p^2}\\
-i\alpha_1\sqrt{\frac{2}{\pi p}}e^{-\frac{i}{2}\theta_p}\frac{F_1+p}{F_1^2+p^2}\\
-i\alpha_2\sqrt{\frac{2}{\pi p}}e^{-\frac{i}{2}\theta_p}\frac{F_2+p}{F_2^2+p^2}\\
-i\alpha_2\sqrt{\frac{2}{\pi p}}e^{-\frac{3i}{2}\theta_p}\frac{F_2+p}{F_2^2+p^2}
\end{array}\right),\quad \tilde{\psi}^2_{cl}(p)=\left(\begin{array}{c} 
-i\beta_1\sqrt{\frac{2}{\pi p}}e^{\frac{3i}{2}\theta_p}\frac{F_1+p}{F_1^2+p^2}\\
-i\beta_1\sqrt{\frac{2}{\pi p}}e^{\frac{i}{2}\theta_p}\frac{F_1+p}{F_1^2+p^2}\\
-i\beta_2\sqrt{\frac{2}{\pi p}}e^{\frac{i}{2}\theta_p}\frac{F_2+p}{F_2^2+p^2}\\
-i\beta_2\sqrt{\frac{2}{\pi p}}e^{-\frac{i}{2}\theta_p}\frac{F_2+p}{F_2^2+p^2}
\end{array}\right).
\eea

\end{document}